# Spin and Orbital Splitting in Ferromagnetic Contacted Single Wall Carbon Nanotube Devices

K. Y. Wang [1,*] , A. M. Blackburn[2], H. F. Wang[1], J. Wunderlich[2], D. A. Williams [2]

[1]SKLSM, Institute of Semiconductors, P. O. Box 912, Beijing 100083, P. R. China
[2] Hitachi Cambridge Laboratory, Cambridge CB3 0HE, United Kingdom
*Email: kywang@semi.ac.cn

We observed the coulomb blockade phenomena in ferromagnetic contacting single wall semiconducting carbon nanotube devices. No obvious Coulomb peaks shift was observed with existing only the Zeeman splitting at 4K. Combining with other effects, the ferromagnetic leads prevent the orbital spin states splitting with magnetic field up to 2 Tesla at 4K. With increasing magnetic field further, both positive or negative coulomb peaks shift slopes are observed associating with clockwise and anticlockwise orbital state splitting. The strongly suppressed/enhanced of the conductance has been observed associating with the magnetic field induced orbital states splitting/converging.

_________________________

* E-Mail: kywang@semi.ac.cn

Carbon nanotube has very unique properties for electronics, such as high current density capacity[1] and long spin flipping length[2,3] , which makes it one of the best candidates for substituting Si-based electronic devices or even quantum computing. Fourfold degeneracy due to both the spin degeneracy and orbital degeneracy are expected in non disturbed carbon nanotube devices at zero magnetic fields[4-7]. The latter degeneracy can be pictured as the separate clock- and anticlock-wise electron states spiraling down the tube. This fourfold degeneracy can be lifted by impurity or in a magnetic field with the orbital splitting and spin Zeeman effects. Spin-orbital coupling has also been observed and tuned by using electrical field in single wall carbon nanotube devices [8,9]. If we can tune the spin and orbital quantum states in single wall carbon nanotube (SWCNT) devices, the different spin or orbital quantum states could be used for information manipulation in quantumn computation. In this letter, we use the electrical gate and magnetic field to tune the quantumn states in ferromagnetic contacted SWCNT devices.

Different type of ferromagnetic metals have been used to study ferromagnetic contacted CNT devices[10-13], the spin polarization of the leads will lift the spin degeneracy of the system even at zero magnetic field and we only expect the orbital degeneracy. Because of the different work function between the carbon nanotube and the ferromagnetic leads, a Schottky barrier is normally formed at the interface between the ferromagnetic metals and carbon nanotubes. In order to create the more reliable interface, we introduce a very thin insulator layer between ferromagnetic lead and carbon nanotube. In this letter, we investigate the Zeeman splitting and orbital splitting of ferromagnetic contacted SWCNT field effect devices with varying the external magnetic field. Due to the thermal fluctuations, the Zeeman splitting can not be resolved at 4 K with magnetic field perpendicular to the substrate. The ferromagnetic contacted device can prevent the orbital splitting with magnetic field along the tube up to 2 Tesla. However, with increasing magnetic field further, both positive and negative coulomb peaks shift slopes are observed at certain back gate voltages associating with clockwise and anticlockwise orbital state splitting.

We pre-patterned the catalyst mark using optical lithography on a $SiO_2/Si^{++}$ substrate, and then the CNTs were grown by chemical vapour deposition. The contact resist mark was fabricated by electron beam lithography, and lift off technology. A 1-nm thick layer of Al was deposited under ultrahigh vacuum and then fully oxidized to $Al_2O_3$ under 100 mbar of pure $O_2$ pressure for 1 h. We then sent the wafer back into the electron beam deposition chamber and pump to ultrahigh vacuum again. 50nm thick 200 and 1000 nm wide Ni stripes acting as source and drain electrodes were deposited on top of the $Al_2O_3$. The schematic diagram of the device is shown in figure 1 a and the scanning electron microscopy (SEM) picture of the device is shown in figure 1 b, where the angle θ between the ferromagnetic bond-pad and the carbon nanotube is about $30^0$. The electrical and magnetotransport measurements were performed using Oxford Heliox system.

Figure 1c shows the measured source drain current $I_{SD}$ dependence of back gate $V_g$ at various source drain voltage $V_{SD}$ at 4 K. With varying the gate voltage from negative to positive, the device can be tuned from p type to n type semiconductor through off state, indicating the bipolar property of the carbon nanotube. The ratio of fully switched on p state and off state current linearly increases with increasing source drain voltage, which achieves $2 \times 10^3$ for $V_{SD}$ =50 mV. This value is rather small because of the tunnelling barrier between the dots and the ferromagnetic leads. The current oscillates with varying the back gate at the weak conducting region, indicating the presence of the coulomb blockade features. To understand the detail physics of this region, plot of the differential conductance dI/dV as a function of bias voltage and gate voltage at 4 K is shown in figure 2, where a series of coulomb diamonds are observed. The analysis of the coulomb diamonds allows us to extract the coupling capacitance efficiency α between the gate and CNT, which measures i.e., $\alpha = C_g / C_\Sigma = U_C / \Delta V_g$, where total capacitance $C_\Sigma = C_S + C_D + C_g$ (the sum of the capacitances between the dot and the source $C_S$, the drain $C_D$, and the gate $C_g$), $U_C = e^2 / C_\Sigma$ the charge energy, and $\Delta V_g$ the single-electron period in gate voltage. Rather large charge energy is estimated to be about 30 meV from the averaged value of the adding energy of a set of Coulomb diamonds, which is corresponding to the gate voltage period of 52 mV. The gate coefficient $\alpha = 0.56$ and total capacitance $C_\Sigma$ = 0.53 fF are obtained. Then a rather large gate capacitances are expected, which is much larger than the value 8.6 aF obtained from $C_g \approx 2\pi\varepsilon\varepsilon_0 L / Ln(4h/D)$,[14] where L the length of the CNT, and $h$ the thickness of the SiO$_2$ layer. From the fully depleted gap region at low source drain voltage in Figure 1c and the gate coefficient, the bandgap of the carbon nanotube is obtained to be ~0.3 eV. The bandgap of semiconducting nanotube is also predicted from the band theory to be $E_g \sim 0.7 eV / d(nm)$[15], where $d$ is the diameter of the carbon nanotube. According to this analysis, the diameter of the single wall CNT is ~2.3 nm.

When the magnetic flux pass through the electron/hole orbital circles, namely along the carbon nanotube, it will break the orbital degeneracy. While with the magnetic field B perpendicular to the substrate plane, only Zeeman splitting $g\mu_B S_\pm B$ and no orbital splitting appears to the carbon nanotube, where $\mu_B$ the Bohr magneton and $S_\pm = \pm 1/2$ for spin up and spin down states respectively. Increasing perpendicular magnetic field from 0 to 9T at 4.2 K, the colour plot of electrical current as a function of external magnetic field and gate voltage at fixed low bias V$_{SD}$ = 5 mV is shown in Figure 3 a. No obvious coulomb peak shift is observed with increasing the external magnetic field at any back gates. We attribute this to the comparable of the thermal fluctuation broadening energy and the Zeeman splitting energy is ± 57.9 μeVT$^{-1}$ with $g = 2$, which limits the resolution of the Zeeman splitting energy.

Two dimensional plot of current versus gate voltage and external magnetic field with field in the plane perpendicular to the magnetic bond-pad at a constant bias voltage V$_{SD}$ = 5 mV is shown in figure 3b and 3c. Compared with the field perpendicular to the carbon nanotube, very distinct behaviour was observed. The coulomb peaks are fixed at relative low magnetic field. Then a kinck was observed with magnetic field increasing further. Both the peak positions and the magnitude of the coulomb peaks vary with the external magnetic field above 2 Tesla. In addition to the Zeeman splitting, the orbital splitting will appear when the carbon nanotube lies in the magnetic field direction. The shifted energy of the electrons due to the appearance of orbital magnetic moment is: $\Delta E = -B_{\parallel}\mu = \pm de\, v_F B\cos\theta / 4$, where $v_F$ the Fermi velocity, $B\cos\theta$ the magnetic field projection along the carbon nanotube, $\pm$ the clockwise and anticlockwise orbital, respectively[16]. The largest Fermi velocity $v_F = 9.\times 10^5\, m/s$ is found to be at the band gap edge[17]. With increasing external magnetic field parallel to the carbon nanotube $B_{\parallel}\cos\theta$, the band gap for clockwise orbital sub-band is increased and the anticlockwise one is decreased. Linearly shifted peak positions are expected with increasing the parallel magnetic field and the shifting directions dependent on the different orbital states[16]. The level crossing will appear when the relative varied energy of two anti-orbital states with increasing the magnetic field is larger than that of the orbital space energy. However, as shown in figure 3b, the coulomb peak was fixed at low magnetic field, followed by a kinck, and then linearly shifted with increasing magnetic field further. The low field coulomb peak fixed is not consistent with our expectations. The orbital states partially pinned has also been observed in previous carbon nanotube devices contacting with normal metal, where they attribute the low field unexpected behaviour to the exchange coupling.[16] The only difference is that the partially pinned orbital states persist to higher external magnetic field in our device, which could be attributed to the ferromagnetic contacts. The magnetic easy axis is determined by the shape anisotropy of polycrystalline ferromagnetic Ni, which lies in the plane along the stripe direction. The magnetic shape anisotropy of Ni stripe can be estimated from $-\mu_0 M_S N$, where $\mu_0$ the permeability of vacuum, $M_S$ the saturation magnetization and $N$ the demagnetization factor. The magnetization anisotropy of our long Ni stripe is 0.6 Tesla with $N = 1$. An external magnetic field (> 0.6 Tesla) is needed to pull the magnetization of the lead fully along the magnetic hard axis direction. Since the orbital moment of the electron in the CNT is in discrete levels, the discrete energy of orbital states can be coupled with the pinned spins and thus the energy. Therefore, the ferromagnetic leads could directly or through the spins in the nanotube couple with the orbital states. Combining with the other effects such as exchange coupling[18], which result in a much smaller slope at low magnetic field region shown in figure 3 b and 3c. Considered the misalignment of the carbon nanotube and the external magnetic field, the energy shifting of the Coulomb peak is 0.52 meV/T from the high field slope. The diameter of the carbon nanotube is estimated to be $d \sim 2.3$ nm from the magnetic field induced orbital splitting. This value is in good agreement with the value obtained from the bandgap.

Both large negative and positive magnetoresistance are observed associating with splitting and merging of orbital states. In figure 3 b, the degenerated clock- and anticlock-wise orbital states start to split at magnetic field above 2 T with $V_g \sim$ -0.96 V, which split into two coulomb peaks with opposite slope. Associating with the shift of the orbital states, the conductance is strongly suppressed associating with the orbital energy shifts. As shown in figure 3c, the neighbour opposite orbital states join together under external magnetic field at certain gating voltages, which result in the strongly enhanced conductance. The differnet high and low resistance quantum states could be used in quantum information, which can be manipulated by using both the electrical gate and the external magnetic field.

In summary, the coulomb blockade behavior has been observed in ferromagnetic contacted semiconducting single wall carbon nanotube field effect device. No obvious coulomb peak shift with magnetic field perpendicular the substrate was observed at 4K due to the thermal energy broadening. With applying magnetic field along the nanotube, the orbital states are pinned up to 2 Tesla at 4K. With increasing magnetic field further, positive and negative coulomb peaks shift slopes appeared associating with clockwise and anticlockwise orbital states. Both large negative and positive magnetoresistance are observed associating with splitting and merging of orbital states. It allows us to tune the different quantum states in a single CNT device by means of manipulation of orbital splitting.

**Acknowledgements:**

This project was supported by"973 Program" No. 2011CB922200, EU "CANAPE", NSFC Grant 10974196 and 61225021. KYW acknowledges support of Chinese Academy of Sciences "100 talent program". KYW also would like to thank P. E. Lindelof and H. I. Jørgensen for collaborations.

**Figure Captions:**

Figure 1:   (a) The schematic structure of the Ni/AlOx/CNT/AlOx/Ni device; (b) the SEM picture of the device; (c) the current versus back gate voltage at different source drain voltages.

Figure2: Two dimensional color plot of the differential conductance dI/dV as a function of bias voltage and gate voltage at 4 K

Figure3: (a) The colour plot of electrical current as a function of gate voltage and external magnetic field perpendicular to the substrate at fixed low bias $V_{SD}$ = 5 mV ;(b, c) With magnetic field in plane perpendicular to the stripes of the ferromagnetic contacts, the colour plot of electrical current as a function of external magnetic field at different gate voltage region at fixed low bias $V_{SD}$ = 5 mV. The white dash lines guide the eye.

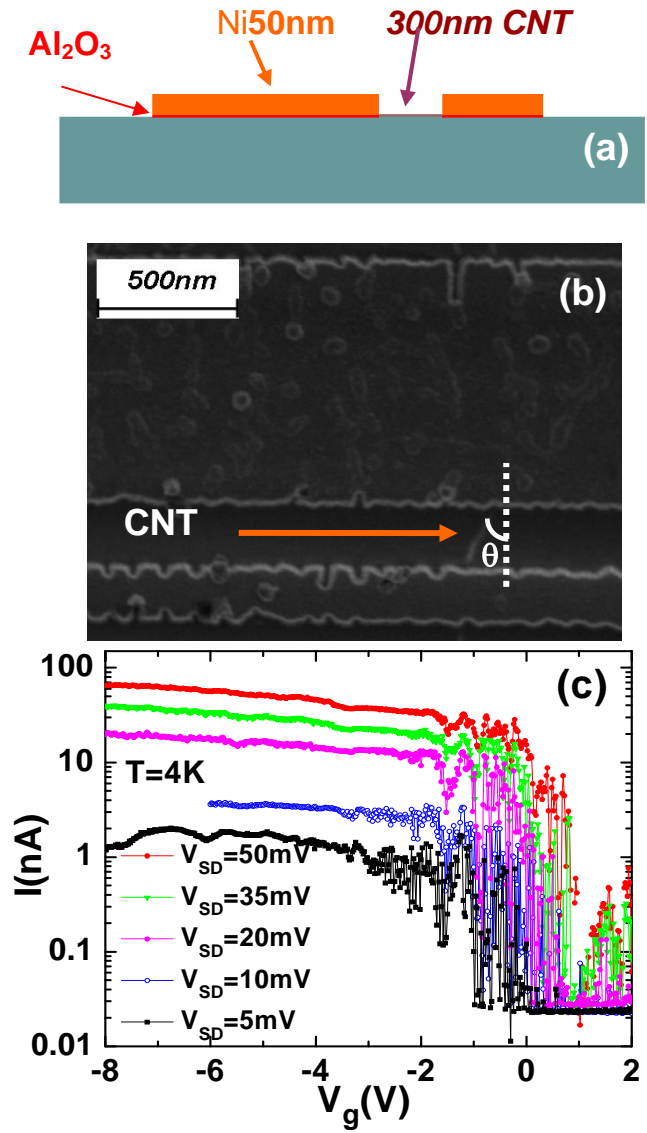

Figure 1 K. Y. Wang et al.

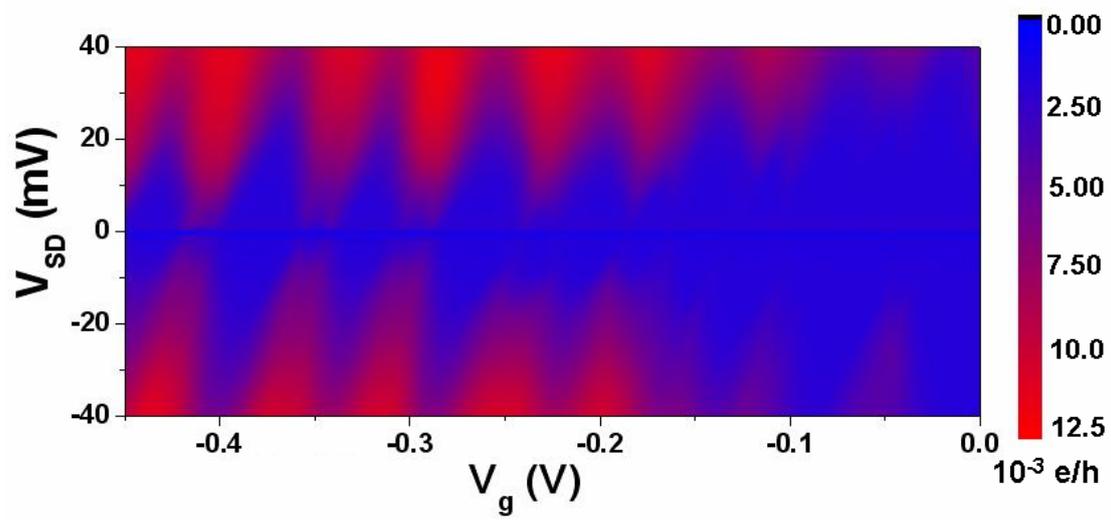

*Figure 2 K. Y. Wang et al.*

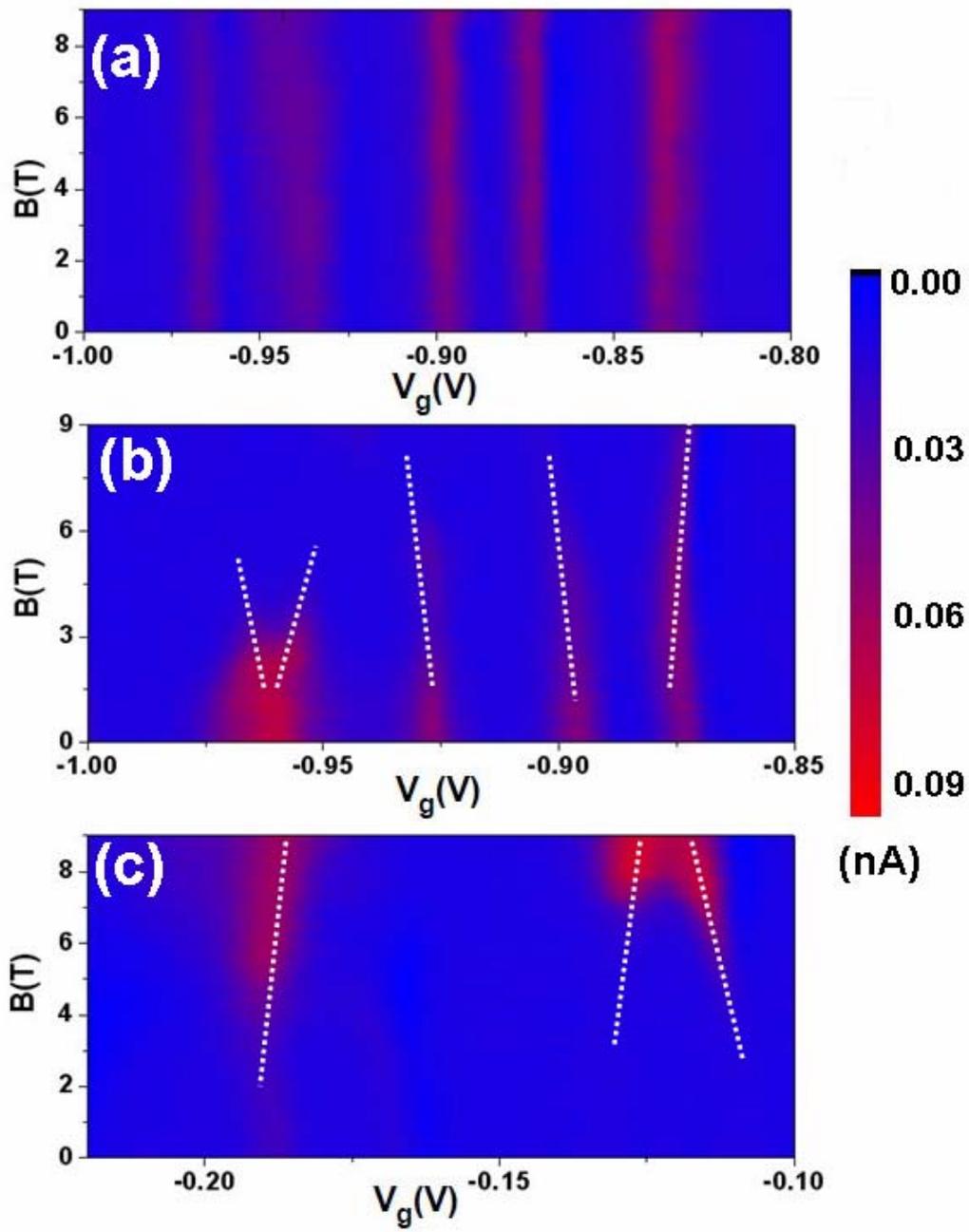

*Figure 3 K. Y. Wang et al.*